\documentclass[letterpaper,aps,prl,superscriptaddress,showpacs,floatfix,twocolumn]{revtex4}
\usepackage{graphicx}
\usepackage{dcolumn}
\usepackage{bm}
\usepackage{xcolor}
\usepackage{color}
\usepackage{soul}

\DeclareRobustCommand{\rchi}{{\mathpalette\irchi\relax}}
\newcommand{\irchi}[2]{\raisebox{\depth}{$#1\tilde{\chi}$}}
\newcommand{\Xuup}{X_u^\uparrow}
\newcommand{\Xudown}{X_u^\downarrow}
\newcommand{\Xdup}{X_d^\uparrow}
\newcommand{\Xddown}{X_d^\downarrow}

\begin{document}
\title{Possible Explanation of $F_2^n/F_2^p$ at Large $x$
Using Quantum Statistical Mechanics}

\affiliation{%
        {Aix Marseille Univ}, Universit\'{e} de Toulon, CNRS, CPT,
Marseille, France}
\affiliation{%
 Istituto Nazionale di Fisica Nucleare, Sezione di Roma 1, 00185 Roma, Italy.}%
\affiliation{Institute of Physics, Academia Sinica, Taipei
11529, Taiwan}
\affiliation{Dipartimento Interateneo di Fisica, Universit\`{a}
degli Studi di Bari Aldo Moro\\Istituto Nazionale di Fisica
Nucleare, Sezione di Bari, 70125 Bari, Italy}
\affiliation{Dipartimento di Matematica e Fisica "Ennio De Giorgi",\\
Universit\`{a} del Salento and INFN-Lecce, Via Arnesano, 73100 Lecce, Italy.}
\affiliation{Department of Physics, University of Illinois at
Urbana-Champaign, Urbana, Illinois 61801, USA}

\author{C. Bourrely}
 \email{claude.bourrely012@orange.fr}
\affiliation{%
	{Aix Marseille Univ}, Universit\'{e} de Toulon, CNRS, CPT,
Marseille, France}
\author{F. Buccella}%
 \email{buccella@roma1.infn.it}
\affiliation{%
 Istituto Nazionale di Fisica Nucleare, Sezione di Roma 1, 00185 Roma, Italy.}%

\author{W. C. Chang}%
 \email{changwc@phys.sinica.edu.tw}
\affiliation{Institute of Physics, Academia Sinica, Taipei 
11529, Taiwan}

\author{D. Di Bari}
 \email{domenico.dibari@uniba.it}
\affiliation{Dipartimento Interateneo di Fisica, Universit\`{a} 
degli Studi di Bari Aldo Moro\\Istituto Nazionale di Fisica 
Nucleare, Sezione di Bari, 70125 Bari, Italy}

\author{P. H. Frampton}%
 \email{paul.h.frampton@gmail.com}
\affiliation{Dipartimento di Matematica e Fisica "Ennio De Giorgi",\\ 
Universit\`{a} del Salento and INFN-Lecce, Via Arnesano, 73100 Lecce, Italy.}

\author{J. C. Peng}%
 \email{jcpeng@illinois.edu}
\affiliation{Department of Physics, University of Illinois at
Urbana-Champaign, Urbana, Illinois 61801, USA}

\date{\today}
\begin{abstract}

The recent accurate measurements of the scattering of electrons off of 
the mirror nuclei $^3$H and $^3$He show with small errors that the 
neutron to proton ratio $r(x)=F^n_2(x)/F^p_2(x)$, approaches a value 
larger than 1/4 for $x \to 1$. 
We suggest a possible explanation for this experimental result by 
studying the consequences of the Pauli exclusion principle for the 
parton distribution when the ratio is described by quantum statistical 
mechanics description in terms of three parameters inspired by the 
quantum statistical approach.

\end{abstract}

\maketitle

\section{Introduction}

Deep Inelastic Scattering (DIS) of electrons from nucleons, and the 
subsequent extraction
of the structure functions, led to the discovery of the 
quarks and antiquarks substructure of the nucleons. It also 
provided a solid basis for establishing
Quantum ChromoDynamics (QCD) as the correct theory for the
strong interaction \cite{SM1,SM2,SM3,SM4,SM5,SM6,SM7,SM8,SM9}.

The differential cross sections for DIS are expressed in terms of the nucleon
structure functions, $F_1$ and $F_2$ :
\begin{equation}
\frac{d^2\sigma}{d\Omega dE^{'}}
= \sigma_M \left[ \frac{F_2(\nu,Q^2)}{\nu} +  2\frac{F_1(\nu,Q^2)}{M} \tan^2
\left( \frac{\theta}{2} \right) \right],
\label{DIS}
\end{equation}
in which $\sigma_M$ is the Mott cross section,
$E^{'}$ and $\theta$ are the energy and polar angle of the scattered electron,
$\nu=E-E^{'}$ where $E$ is the incident electron energy, 
$Q^2= 4EE^{'} \sin^2 \theta/2$ is the negative of the momentum transfer
squared, and $M$ is the nucleon mass.



In the quark-parton model, DIS is represented as
scattering of electrons from point-like constituents, carrying the
momentum fraction, $x$, of the nucleon. In the infinite
momentum scaling limit when $\nu\rightarrow\infty, Q^2\rightarrow\infty$,
$0\leq (x\equiv Q^2/2 M \nu) \leq 1$, the structure function $F_2$ becomes
\begin{equation}
F_2(x) =  x\Sigma_i e_i^2 f_i(x),
\label{F2}
\end{equation}
where $f_i(x) dx$ is the probability that a parton of type $i$ carries
momentum in the range between $x$ and $x+dx$, and the sum in Eq.~(\ref{F2})
runs over all parton types.

We consider the up ($u$), down ($d$), and strange ($s$) quarks in the
proton and define
$U(x) \equiv u(x)+\bar{u}(x)$, $D(x) \equiv d(x)+\bar{d}(x)$ and
$S(x) \equiv s(x)+\bar{s}(x)$. With these notations, together with the
assumption of isospin symmetry for the nucleon PDFs, we have
\begin{equation}
F_2^p(x)=x\left[ \frac{4}{9} U(x) + \frac{1}{9} D(x) + \frac{1}{9} S(x)
\right]
\label{FpUDS}
\end{equation}
\noindent
and
\begin{equation}
F_2^n(x)=x\left[ \frac{1}{9} U(x) + \frac{4}{9} D(x) + \frac{1}{9} S(x)
\right]
\label{FnUDS}
\end{equation}

Because all parton distribution functions are non-negative we deduce from
Eqs.~(\ref{FpUDS}) and (\ref{FnUDS}) that the $F^n_2(x)/F^p_2(x)$
ratio, $r(x)$, is bounded at all $x$ by
\begin{equation}
1/4 \leq \left( r(x) \equiv \frac{F_2^n(x)}{F_2^p(x)} \right)    \leq 4
\label{ratiolimits}
\end{equation}
which is known as the Nachtmann inequality~\cite{Nachtmann}.

The earliest DIS experiments at SLAC~\cite{GottfriedExperiment} 
confirmed the Nachtmann
inequality and, more specifically, discovered that, for $x \rightarrow 0$,
$r$ is approximately unity, whereas for $x\rightarrow 1$,  with
much larger errors, it was thought that $r$ might asymptotically 
approach $1/4$.
However, the uncertainties of the nuclear effects for the deuteron nucleus
have raised some concerns on the validity of the extraction of $r(x)$ at large
$x$~\cite{Whitlow92,Thomas96}. 
New measurements which would minimize the 
nuclear effects were proposed for a more reliable determination of $r(x)$
at large $x$~\cite{Afnan00}.

Recently, the MARATHON experiment~\cite{Abrams}, performed by
the Jefferson Laboratory Hall A Tritium Collaboration, 
has measured the DIS of electrons from the
mirror nuclei $^3$H and $^3$He in order to cleanly separate the proton and
neutron structure functions.
The measurement of $r(x)$ covers a broad range of $x$, from $x = 0.19$ 
up to $x=0.83$. As shown in Fig. 1, the 
$F^n_2(x)/F^p_2(x)$ from the MARATHON experiment exhibits an
intriguing $x$ dependence. At intermediate $x$ region ($0.19 < x < 0.5$),
$F^n_2(x)/F^p_2(x)$ falls roughly linearly with $x$, while for
the large $x$ region, $F^n_2(x)/F^p_2(x)$ falls off more slowly with $x$,
approaching a constant value of $\sim 0.45$ at the highest values of $x$.
\begin{figure}[!ht]
\centering
\includegraphics[width=\columnwidth]{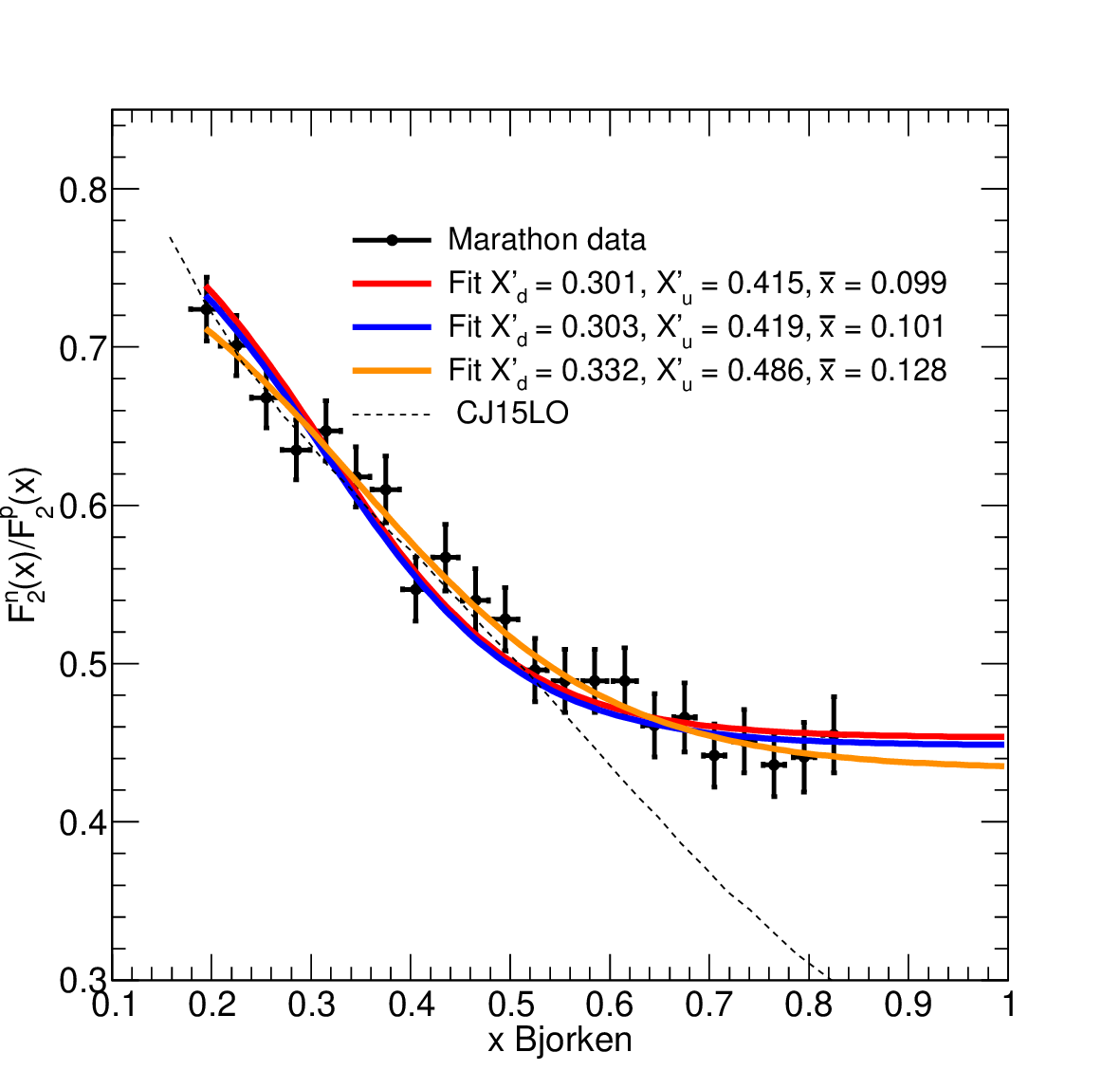}
\caption{Comparison of the MARATHON data with calculations. The dashed
curve is obtained using the CJ15LO proton PDFs~\cite{CJ}, taking into
account the $Q^2$ dependence of the MARATHON data.
The solid curves are obtained from the statistical model using the two
sets of parameters listed in Table I:
$\bar{x}=0.099$ (red), $\bar{x}=0.101$ (blue) and $\bar{x}=0.128$ (orange), respectively.}
\label{fig_Mar}       
\end{figure}

As discussed in the MARATHON paper, the observed $x$ dependence for 
$F^n_2(x)/F^p_2(x)$ is at a variance with the prediction of the
CTEQ-JLab (CJ) Collaboration, based on a global fit of
existing data using the conventional functional forms for describing the
parton distributions~\cite{CJ}. 
As shown by the dashed curve in Fig. 1, obtained by using the CJ15LO proton 
PDFs, falls off with $x$, approaching a value of 1/4 as $x \to 1$, in 
disagreement with the MARATHON data. It should be 
noted that the MARATHON data
were not available when the CJ15LO PDFs were obtained. The MARATHON data
would clearly impact on future global fits to extract the proton PDFs,
especially at the large $x$ region.

The MARATHON results on $F^n_2(x)/F^p_2(x)$ in a broad
region in $x$ up to the largest value $x = 0.83$ have inspired several recent
papers to address various aspects of the 
data~\cite{Va,Ha,Ab,Li,Gr,Ar,Ac,Ar1,Al,Ro}, including the off-shell 
contributions~\cite{Ha} and the EMC effect~\cite{Ab}.

In this paper, we show that the MARATHON data on 
$F^n_2(x)/F^p_2(x)$ can be well described 
by an approach based on the quantum statistical mechanics. The good agreement
between the MARATHON data and the statistical approach lends further
support for the validity of this approach in depicting the partonic structures
of hadrons.

\section{The Quantum Statistical Description of Parton Distributions}

We first briefly discuss the salient features of the quantum 
statistical approach in depicting the dynamics of partons
inside the nucleons.  
The rule of the Pauli Exclusion 
Principle \cite{NS,FF} implies quantum statistical 
parton distributions, namely Fermi-Dirac type for the quarks and 
antiquarks and Bose-Einstein type
for the gluons. In 2002, Bourrely, Buccella and Soffer~\cite{BBS}
proposed the following distributions at the initial scale, 
$Q_0 = 2$ GeV/c for the quarks and antiquarks,
$q(x)$ and $\bar q(x)$, and the gluons, $G(x)$, 
on the basis of quantum statistical mechanics:
\begin{eqnarray}
x q^h (x) &=& A X_q^h x^b  \left[\exp\frac{(x-X_q^h)}{\bar{x}} + 1
\right]^{-1} \nonumber\\
 &+& \tilde{A} x^{\tilde{b}}\left[\exp\frac{x}{\bar{x }} +1 \right]^{-1}
\label{QSMq}
\end{eqnarray}
\noindent
and
\begin{eqnarray}
x \bar{q}^h (x) &=&
\bar{A} x^{\bar{b}}
\left[X_q^{-h} (\exp\frac{(x+X_q^{-h})}{\bar{x}} 
+1) \right]^{-1}\nonumber\\
 &+& \tilde{A} x^{\tilde{b}} \left[\exp\frac{x}{\bar{x}} +1 \right]^{-1}
\label{QSMqbar}
\end{eqnarray}
\noindent
and
\begin{equation}
x G (x) =  A_G x^{b_G}\left[\exp \frac{(x-X_G)}{\bar{x}} -1  \right]^{-1}.
\label{QSMg}
\end{equation}

In these distributions, $\bar{x}$ plays the r\^{o}le of temperature.
$X_q^h$ are chemical potentials depending on the flavour ($q$ = $u$ or $d$)
and the helicity ($h$). The factors $X_q^h$ in Eq.~(\ref{QSMq}) 
and  $(X_q^{-h})^{-1}$
in Eq.~(\ref{QSMqbar}) were introduced in \cite{BBS}
to comply with the data and have been accounted for by
considering the transverse degrees of freedom~\cite{BBS4}.
The normalization factors $A,\tilde{A},\bar{A}$ and the exponents 
$b,\tilde{b},\bar{b}$ are determined by fitting the data, together with the
constraints of the quark number and the momentum sum rules. 
For the strange partons, it was 
assumed that $s(x)=\bar{s}(x) = (\bar{u}(x)+\bar{d}(x))/4$.

The Fermi-Dirac form for the quark and antiquark distributions are very
different from the $Ax^B(1-x)^C$ form for standard parametrizations. This
difference would lead naturally to different behaviors of $F^n_2(x)/F^p_2(x)$
between the statistical model and the conventional parametrization.

The equilibrium conditions with respect to the processes which lead to the
DGLAP equations~\cite{DGLAP,DGLAP1} imply that 
\begin{equation}
X_q^h + X_{\bar{q}}^{-h} = 0
\label{Qpot}
\end{equation}
and
\begin{equation}
X_G = 0.
\end{equation}
Equation (\ref{Qpot}) provides a natural connection between the valence and sea
quark parton distributions, since the chemical potentials of the quark $q$
and antiquark $\bar q$ are related. It also implies that the helicities
of the quark and antiquark are correlated. These intriguing correlations
between quark and antiquark distributions, and between their positive and 
negative helicity distributions,
are unique features of the quantum statistical approach and they are
absent in the usual standard parametrizations of nucleon PDFs in the 
conventional global fits. 

The values of $\bar{x}$ and $X^h_q$, found in the first 
paper~\cite{BBS} in 2002, are respectively:
\begin{eqnarray}
\bar{x}=0.099; ~ \Xuup &=& 0.461; ~ \Xddown=0.302;
\nonumber\\
\Xudown &=&0.298; ~\Xdup=0.228.
\label{values1}
\end{eqnarray}
Following the cited paper~\cite{BBS}, supporting confirmations were
forthcoming for the quantum statistical parton distributions proposed
therein. The temperature $\bar{x}$ and the quark chemical potentials were obtained,
for example, in the 2015 global fit (BS15) in the quantum statistical 
approach~\cite{BS1} with $Q_0 = 1$ GeV, as follows:
\begin{eqnarray}
\bar{x}=0.090; ~ \Xuup &=& 0.475; ~ \Xddown=0.309;
\nonumber\\
\Xudown &=&0.307; ~\Xdup=0.245.
\label{values2}
\end{eqnarray}
The values of the parameters do not need to be equal,
since the DGLAP equations \cite{DGLAP,DGLAP1} imply that the parton distributions depend on x.
One observes the following relations for the chemical 
potentials of the valence quarks:
\begin{equation}
\Xuup > \Xddown \approx \Xudown > \Xdup.
\label{valence}
\end{equation}
Eqs. (\ref{Qpot}) and (\ref{valence}) imply the following
relations for the chemical potentials of the antiquarks:
\begin{equation}
	X_{\bar{u}}^{\downarrow} < X_{\bar{d}}^{\uparrow} \approx  
	X_{\bar{u}}^{\uparrow} <X_{\bar{d}}^{\downarrow}.
\label{inequalitiesqbar}
\end{equation}
Eq. (\ref{inequalitiesqbar}) leads to the following 
striking predictions of the quantum statistical approach for 
the flavor and spin structure of the 
antiquarks in the proton: 
\begin{equation}
\bar{d}(x) > \bar{u} (x)
\label{INEQ1}
\end{equation}
and
\begin{equation}
\Delta \bar{u}(x)  > 0 > \Delta \bar{d} (x)
\label{INEQ2}
\end{equation}
and finally
\begin{equation}
\Delta \bar{u} (x) - \Delta \bar{d}(x) \approx \bar{d} (x) - \bar{u} (x).
\label{INEQ3}
\end{equation}

The first inequality, Eq. (\ref{INEQ1}), has been confirmed by the
Fermilab E866 experiment~\cite{E866a,E866b,E866c}
and the SeaQuest experiment~\cite{Do1,Do}. 
The second inequality, Eq.~(\ref{INEQ2}), has been
confirmed by the STAR Collaboration at RHIC on 
the production of charged weak bosons
using polarized beams~\cite{JAdam} .
The test of the last relation, Eq.~(\ref{INEQ3}), still awaits a 
more precise determination of the quantity on the left-hand side.

The gluon parton distribution proposed by ATLAS has been described by the
three parameter Planck formula proposed in \cite{BBS}, with the same
value of $\bar{x} =0.099$ and values of the other two parameters
$A_G$ and $b_G$ similar to those found therein; see
\cite{Bellantuono2022}.

An analysis of DIS at  HERA with the statistical parametrization could
describe the data with less parameters and a similar $\rchi^2$ to that
obtained with the standard parametrization, with a good agreement of the
non-singlet distributions of $u(x)$ and $d(x)$ with a value of
$\bar{x}=0.097$ \cite{Bonvini2023}.
Very recently, quantum statistical parton distributions have been successfully
applied to pions \cite{BBP,BCP} and kaons \cite{BBCP}
with values of $\bar{x}$ around $0.1$.

\section{The Quantum Statistical Description of $F_2^n(x)/F_2^p(x)$}

We now turn to the implications of the quantum statistical approach for 
understanding the behavior of the MARATHON $F^n_2(x)/F^p_2(x)$
data. As shown in Fig.~\ref{fig_Mar}, the ratio $F^n_2(x)/F^p_2(x)$
decreases with a positive curvature, approaching a constant value 
of $\sim0.45$ which is significantly greater than the value of $1/4$ 
expected for $d(x)/u(x) =0$. 

The conventional parametrization for the proton quark distributions 
such as the CJ15LO~\cite{CJ} PDFs, 
often involves a form $(1-x)^{C_q}$ for quark $q$
as $x \to 1$. The ratio $d(x)/u(x)$ becomes proportional to
$(1-x)^{C_d-C_u}$ which goes to zero, since $C_d > C_u$ as one
expects more $u(x)$ than $d(x)$. As $x \to 1$, this conventional 
parametrization naturally leads to $d(x)/u(x) =0$ and 
$F^n_2(x)/F^p_2(x) \to 1/4$. Modifications to the parametrization
of $(1-x)^{C_q}$ have been introduced to allow more 
flexibility~\cite{CJ}. As shown in Fig.~\ref{fig_Mar}, the MARATHON data
would provide further constraints for determining the functional form for
$q(x)$ in future global fits. In contrast to the conventional approach
allowing for flexible function forms for $q(x)$, the statistical 
model requires a Fermi-Dirac form for the quark distribution which dictates
the shape of $q(x)$ at large $x$. As discussed next,
the Fermi-Dirac form for the quark distributions in the quantum statistical
approach would lead to a very different behavior for
$d(x)/u(x)$ as $x \to 1$.

We note that at large $x$ the valence quarks dominate 
and therefore we may write :

\begin{equation} 
\frac{F^n_2(x)}{F^p_2(x)} = \frac{4 d(x) + u(x)}{4 u(x) + d(x)}
\label{VD}
\end{equation}

\noindent
The sum of two Fermi-Dirac functions is well approximated by
a single Fermi-Dirac function with a potential intermediate,
which allows us to write :

\begin{equation}
x q(x) = \frac{A'_1 X'_q x^b}{\exp{\frac{x - X'_q}{\bar{x}}} + 1}
\label{q(x)}
\end{equation}

\noindent
with

\begin{equation}
X'_u = \Xudown +
\bar{x} \ln{\frac{[\Xuup
\exp{\frac{\Xuup - \Xudown}{\bar{x}}
+ \Xudown}]}
{[\Xuup + \Xudown]}}
\label{X'_u}
\end{equation}

\begin{equation}
X'_d = \Xdup +
\bar{x} \ln{\frac{[\Xddown
\exp{\frac{\Xddown - \Xdup}{\bar{x}}
+ \Xdup}]}
{[\Xdup + \Xddown]}}
\label{X'_d}
\end{equation}

Taking into account Eq. (\ref{X'_u}), for example, for $\Xudown=\Xuup$ or 0
it gives $X'_u= \Xuup$; if $0<\Xudown<\Xuup$  Eq. (\ref{X'_u}) gives 
$X'_u$ in the upper part of the range [$\Xudown, \Xuup$]. The same 
consideration applies to $X'_{d}$ by using Eq. (\ref{X'_d}).
%
Equation (\ref{q(x)}) implies
\begin{equation}
\frac{d(x)}{u(x)}=
\frac{X'_d}{X'_u}\frac{\exp{\frac{x-X'_u}{\bar{x}}} +1}
{\exp{\frac{x-X'_d}{\bar{x}}} +1}.
\label{du(x)}
\end{equation}
The MARATHON data are then fitted according 
to Eq. (\ref{VD}) and Eq. (\ref{du(x)}).
%

\begin{figure}[!ht]
\centering
\includegraphics[width=\columnwidth]{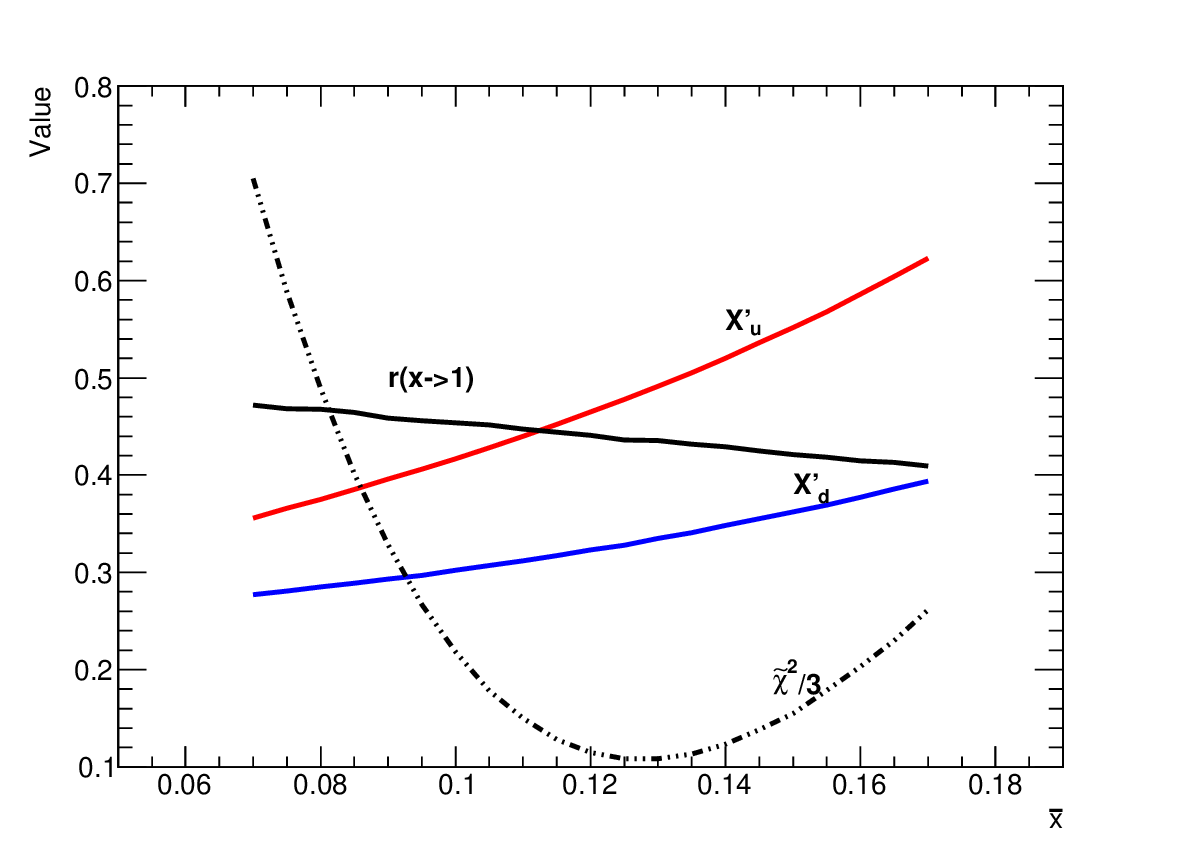}
\caption{Normalized chi square $\rchi^2$ (dashed black), 
$X'_d$ (blue), $X'_u$ (red), and $r(x\rightarrow1)$ (black)
as a function of $\bar{x}$.}
\label{fig_chi2}
\end{figure}
In Fig.~\ref{fig_chi2}, the mean $\rchi^2$ as a function of
$\bar{x}$ is shown; it reaches its minimum at $\bar{x} = 0.128$, which
is somewhat larger than the previously determined value 
of $\bar{x}$ (see Eq. (\ref{valence}), where the parameters correspond to the upgraded fit at $Q_0 = 1~$GeV$/c$) .
Table~\ref{tab:table1} lists  the values 
of $X'_u$, $X'_d$ and $r(x\rightarrow1)$, 
respectively, obtained from the fit to the MARATHON data
using Eq. (\ref{VD}) and Eq. (\ref{du(x)}) for three 
different values of the parameter $\bar{x}$. The first $\bar x$ is from Eq.~(\ref{values1}), 
the second one from Eq.~(\ref{values3}) obtained from a new global fit to be discussed later, while the third one corresponds to the location
of minimal $\rchi^2$ in Fig.~\ref{fig_chi2}.

The values reported in Table~\ref{tab:table1} for $X'_u$ and $X'_d$ may be compared with the ones given by Eqs. (20-21), 0.4235 and 0.2763 for $\bar{x} =  0.099$, 0.4055 and 0.2791 for $\bar{x} = 0.101$, in good agreement for $X'_u$ and with a slight tension for $X'_d$.

The comparison with the MARATHON data for the three cases of $\bar x = 0.099$, $\bar x = 0.101$ and $x = 0.128$ is shown
in Fig.~\ref{fig_Mar}. Excellent agreement between the data and the calculation
are found for the the values of 0.99 and 0.101 for $\bar{x}$, suggesting the relative insensitivity
of $F^n_2(x)/F^p_2(x)$ to the exact values of the parameters adopted in the
quantum statistical approach.

It is instructive to examine how the $F^n_2(x)/F^p_2(x)$ ratios vary for 
different ranges of $x$. The data from MARATHON, as shown in
Table 3 of~\cite{Abrams2}, can be divided into two different $x$ ranges,
[0.195, 0.51] and [0.51, 0.825]. One sees that the $F^n_2(x)/F^p_2(x)$
ratio decreases by $0.212 \pm 0.026$
in the first range and only by $0.057 \pm
0.019$ in the second, in good agreement with the quantum statistical 
approach.

\begin{figure}[!ht]
\centering
\includegraphics[width=\columnwidth]{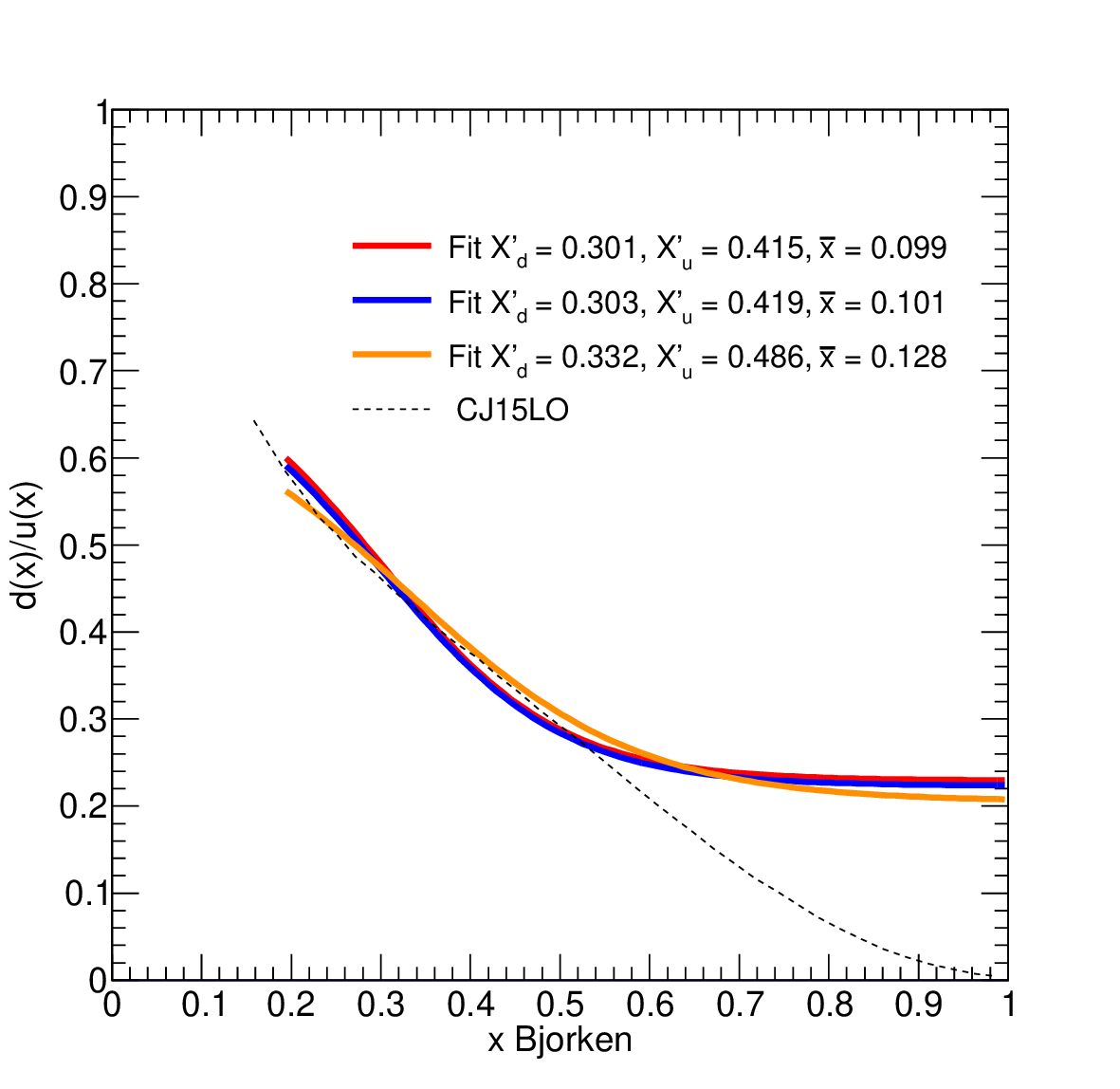}
\caption{The $x$ dependence of $d(x)/u(x)$ for three proton PDFs. The dashed
curve corresponds to the CJ15LO proton PDFs~\cite{CJ}.
The solid curves correspond to the proton PDFs obtained from the 
statistical model using the three sets of parameters listed in Table I:
$\bar{x}=0.099$ (red), $\bar{x}=0.101$ (blue) and
$\bar{x}=0.128$ (orange), respectively.}
\label{fig_dovu}       
\end{figure}
\begin{table}[h]
\caption{\label{tab:table1}
Evaluation of $X'_u$, $X'_d$ and $r(x)$ at $x \rightarrow1$, 
respectively, from the fit to the MARATHON data
using Eq. (\ref{VD}) and Eq. (\ref{du(x)}) for three different 
values of the parameter $\bar{x}$. In the last column are the values of the 
normalized $\rchi^2$.}
\begin{ruledtabular}
\begin{tabular}{c cccc}
\textrm{$\bar{x}$}&
\textrm{$X'_u$}&
\textrm{$X'_d$}&
\textrm{$r(x\rightarrow1)$}&
\textrm{$\rchi^2$}\\
\colrule
\hline
0.099 & 0.415$\pm$ 0.013 & 0.301$\pm$ 0.010 & 0.455$\pm$ 0.006 & 0.68 \\

0.101 & 0.419$\pm$ 0.013 & 0.303$\pm$ 0.010 & 0.453$\pm$ 0.005 & 0.63
\\

0.128 & 0.486$\pm$ 0.018 & 0.332$\pm$ 0.013 & 0.436$\pm$ 0.007 & 0.32 \\
\end{tabular}
\end{ruledtabular}
\end{table}
We also compare the $x$ dependence of $d(x)/u(x)$ with the proton PDFs
obtained in the conventional parametrization and in the quantum statistical
approach in Fig. 3. The dashed curve obtained with the CJ15LO proton 
PDFs~\cite{CJ} falls rapidly with $x$, approaching 0 as $x \to 1$. In contrast,
the proton PDFs obtained from the quantum statistical approach, shown as the 
solid curves corresponding to the two sets of parameters in Table I, have
a much slower fall off at large $x$. The distinct behavior of the $x$ 
dependence of $d(x)/u(x)$ is a result of the Fermi-Dirac form of the quark
distribution in the quantum statistical approach.

While the MARATHON data can be well described by the statistical model
using Eq. (\ref{VD}) and Eq. (\ref{du(x)}), it
is important to verify that the inclusion of the MARATHON data does not
deteriorate the overall quality of fits to other existing data. 
To this end, we have
performed a global fit with the statistical model. This new global fit
includes all previous data contained in the 2015 global 
fit, BS15~\cite{BS1}, plus the MARATHON data and the latest
SeaQuest Drell-Yan data~\cite{Do1,Do}. A total of 2208 data points are
included in this new global fit. Like BS15, this new global fit uses
$Q_0 = 1$ GeV and the PDFs are evoluted to higher $Q$ corresponding to
the experimental data using NLO QCD. From this new global fit the temperature
$\bar x$ and the quark chemical potentials are obtained as follows:
\begin{eqnarray}
\bar{x}=0.101; ~ \Xuup &=& 0.438; ~ \Xddown=0.302;
\nonumber\\
\Xudown &=&0.337; ~\Xdup=0.240.
\label{values3}
\end{eqnarray}
Figure 4 shows that the MARATHON data can be well fitted by this new global 
fit in the statistical model. More details about this new global fit, which 
shows that a simultaneous fit to the existing data together with the MARATHON
data can be obtained in the statistical model, will be
presented elsewhere~\cite{Bourr}.

\begin{figure}[!ht]
\centering
\includegraphics[width=\columnwidth]{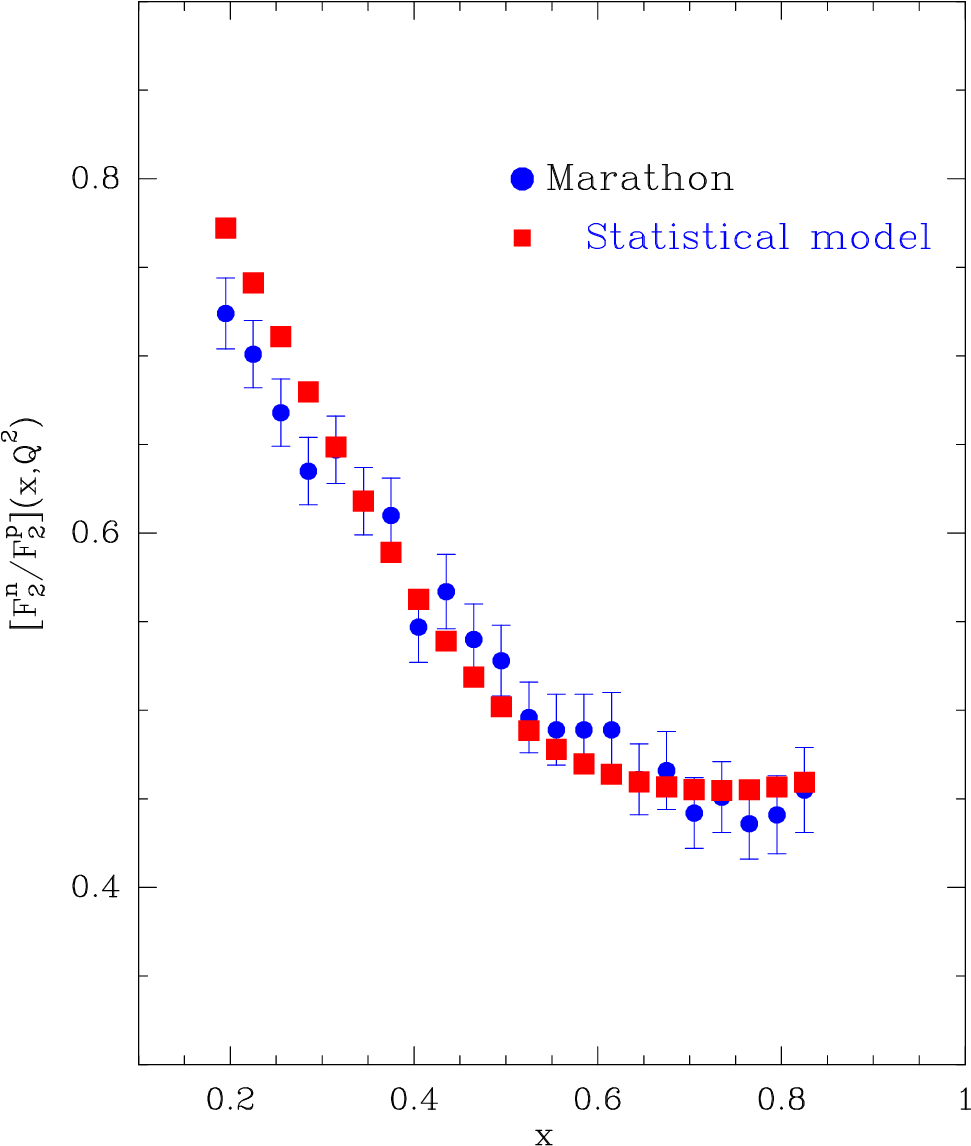}
\caption{Comparison of the MARATHON data with the calculation using
the latest global fit in the statistical model described in the text.}
\label{fig_Claude}       
\end{figure}

\section{Conclusion}

The ratio $F^n_2(x)/F^p_2(x)$, measured
with high precision by the MARATHON experiment, is well described by quantum 
statistical approach, which predicts that the ratio approaches a
constant value greater than 1/4. 
The fact that the ratio decreases faster in the lower region of $x$ 
than in the higher region is also a property of the quantum 
statistical approach.

Unlike conventional parametrizations for the nucleon PDFs, the
statistical approach has imposed specific forms for the parton distributions
based on the Fermi-Dirac nature of the quarks and the Bose-Einstein
nature of the gluons. Various predictions of the statistical approach
are found to be in excellent agreement with existing data. 
Further stringent tests of the quantum statistical approach 
could be performed by considering the
$Q^2$ dependence of the $F^n_2(x)/F^p_2(x)$ ratios, 
as well as the polarized quark distributions. 

It could be, finally, stressed that the distributions of the valence
quarks at the large $x$ region show a big difference with the standard 
distributions. 
Our discussion of using quantum statistical mechanics for the proton 
PDFs could have implications for proton-proton collisions at the LHC.

\end{document}